\documentclass{elsart}

\usepackage{graphics}
\usepackage{graphicx}
\usepackage{epsfig}

\usepackage{amssymb}

\begin{document}

	\begin{frontmatter}



\title{Pion emission in $^2$H,$^{12}$C,$^{27}$Al($\gamma$,$\pi^+$) reactions at 
threshold}

\author[1]{P. Golubev},
\author[1]{V. Avdeichikov}, 
\author[1]{K.G. Fissum}, 
\author[1]{B. Jakobsson\corauthref{cor1}},
\author[2,3]{I. A. Pshenichnov},
\author[4]{W. J. Briscoe}, 
\author[5]{G. V. O'Rielly},
\author[6]{J. Annand}, 
\author[7]{K. Hansen},
\author[1,7]{L. Isaksson},
\author[8]{H. J\"{a}derstr\"{o}m},
\author[1]{M. Karlsson},
\author[7]{M. Lundin},
\author[1]{B. Schr\"{o}der},  
\author[9]{L. Westerberg}, 

\corauth[cor1]{Corresponding author: Bo.Jakobsson@nuclear.lu.se }

\address[1]{Dept. of Physics, Lund University, Box 118, SE 221 00 Lund, 
Sweden}
\address[2]{Frankfurt Inst. for Advanced Studies, J.W. Goethe 
University, 60438 Frankfurt am Main, Germany}
\address[3]{Inst. for Nuclear Research, Russian Academy of Science, 117312 
Moscow, Russia}\address[4]{The George Washington University, Washington, DC 
20052 USA}\address[5]{University of Massachusetts Dartmouth, North Dartmouth, MA 
02747 USA}\address[6]{Dept. of Physics, University of Glasgow, UK}
\address[7]{MAX-lab, Lund University, Ole R\"{o}mers v\"{a}g 1, Lund, Sweden}
\address[8]{Dept. of Nuclear and Particle Physics, Uppsala University, Box 
535, SE 751 21 Uppsala, Sweden}
\address[9]{Dept. of Physics, Uppsala University, Box 530, SE 751 21 
Uppsala, Sweden}

\begin{abstract}
The very first data from MAX-lab in Lund, Sweden on pion photoproduction at 
threshold energies are presented. The decrease of the total $\pi^+$ yield 
in  $\gamma$ + $^{12}$C, $^{27}$Al reactions below 200 MeV as well as 
the d$\sigma$/d$\Omega$ cross-section data essentially follow the predictions 
of an intranuclear-cascade model with an attractive potential for the pion-nucleus 
interaction. However, d$^2\sigma$/d$\Omega$dT, 
cross-section data at 176 MeV show deviations which call 
for refinements of the model and possibly also for the inclusion of  
coherent pion-production mechanisms.  \end{abstract}

\end{frontmatter}


\section{Introduction}
\label{Intr}

Studies of ($\gamma ,\pi$) reactions on nuclei provide information on the in-medium 
pion-nucleon ($\pi N$) interaction~\cite{Krusche,Hombach}, the properties of excited 
spin-isospin flip states of residual nuclei~\cite{Kobayashi,Shoda} and the 
properties of few-body systems~\cite{Laget}. A proper theoretical description of 
photonuclear reactions and in particular of pion photoproduction on nuclei, is a 
prerequisite for calculations performed for the photodisintegration of ultra-high 
energy nuclei in the cosmic microwave background~\cite{Khan} and on nuclear fragmentation 
reactions induced by virtual photons~\cite{Scheidenberger}.  

Unfortunately, data on nuclear pion photoproduction are scarce in the near-threshold 
region because of lack of relevant photon beams. The recently upgraded nuclear-physics beamline 
at MAX-lab is one of only a handful of facilities worldwide which can now provide 
a photon beam of appropriate energy. This is done by colliding a pulse-stretched electron 
beam~\cite{Adl1,Adl2} with a thin radiator and 
momentum analyzing the post-bremsstrahlung electrons in one 
of two tagging spectrometers ~\cite{Tag}. That said, for the early measurements reported on in 
this paper, the tagging spectrometers were not yet commissioned. Thus, in this untagged 
experiment, the photon-endpoint energy was 189 MeV and the average photon-beam 
energy was estimated to be 176 $\pm$ 2 MeV (see section 2.6). 

Some data on near-threshold pion production exist~\cite{Kobayashi,Shoda,Fissum} but 
these come mainly from experiments with rather high pion detection thresholds. 
In this experiment, the two range telescopes allowed for a relatively low threshold 
for pion detection (8.4 and 8.7 MeV) which was achieved using a 
thin (3 mm) first plastic detector and a compact telescope design. Unfortunately, the 
limited solid-angle coverage and the fact that the absolute normalization of the data 
reported on here utilizes ($\gamma$,$p$) 
cross-sections from the literature result in large statistical and systematic uncertainties.  

The measured total cross-section data for $\pi^+$ photoproduction 
on $^2$H, $^{12}$C and $^{27}$Al, together with the corresponding differential (d$\sigma$/d$\Omega$) 
and double differential (d$^2\sigma$/d$\Omega$dT) cross-section data have been compared 
with results from the RELDIS Monte Carlo model for photonuclear 
reactions~\cite{Iljinov,Pshenich}. In this model, the quasi-deuteron absorption 
mechanism coexists with quasifree meson photoproduction on individual 
nucleons. 

\section{Experimental details}
\label{Expd}
\subsection{Photon beam and targets}

The stretched electron beam from the MAX I ring~\cite{Adl1} has long been used for 
photonuclear experiments at the Tagged-Photon Facility at MAX-lab. As previously mentioned, 
in this, very first ($\gamma,\pi$) experiment, the tagging spectrometers~\cite{Adl2} 
were not in operation. The electron beam was simply passing through 
a 150 $\mu$m thick Al radiator which produced a photon beam with bremsstrahlung 
energy distribution in the interval 0 - 189 MeV. These photons 
passed through a 19 mm diameter collimator before impinging on experimental targets of 
C, CD$_2$ and Al with thickness 1 - 2 mm. All results on $\gamma$ + $^2$H reactions have 
been determined from the data on C and CD$_2$ targets.  

\subsection{Pion detectors}

The technique used to identify $\pi^+$ in this experiment was based 
on measuring the 26 ns $\pi\mu$ decay in plastic scintillator range telescopes (Fig. 1). 
The efficiency for identifying pions depends here mainly - but not solely - on 
telescope geometry. It was important to minimize both the electronic noise through 
proper grounding  and the room background through concrete shielding. 
After such precautions, we found that plastic scintillator telescopes of the CHIC 
design~\cite{Ber} were very effective at selecting pions.
\begin{figure}
\begin{center}
\includegraphics[angle=0, width=0.8\textwidth]{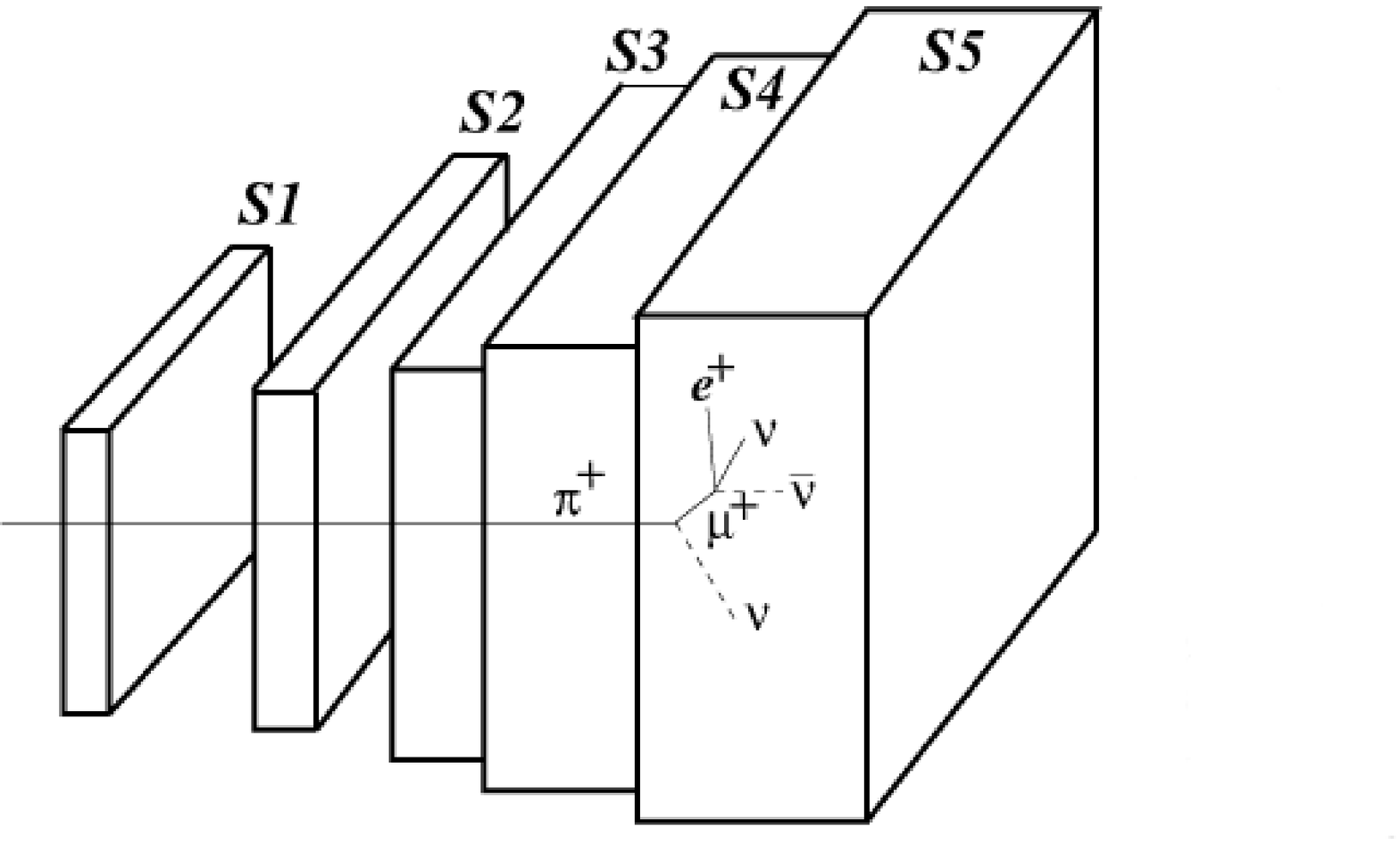}

\caption{Principle geometry and $\pi^+$ detection process in the range 
telescopes (not to scale).} \end{center}
\vspace{0.3in}
\end{figure}

Consequently, two telescopes of this type were designed for the  
experiment under the assumption of an effective upper limit in pion energy 
well below 50 MeV. Each telescope was comprised of five NE102 plastic scintillators 
(S$_1$ - S$_5$), which were read out by Philips XP2020 photomultipliers (PMTs). Table 1 
presents the geometries chosen for the telescopes and the resulting energy bins for protons 
and pions that stop in each of the detector elements. Since pion identification required 
at least two energy signals (see below), the low-energy limit for pion detection was 
set essentially by the thickness of element S$_1$. Including the minor 
energy loss that pions experience in target material, air and wrapping of the scintillators 
(which is different for the two telescopes), the detection thresholds for pions were 
8.4 $\pm$ 0.1 MeV and 8.7 $\pm$ 0.1 MeV for the 30$^\circ$ and 90$^\circ$ 
telescopes respectively. The maximum pion energies each telescope could stop were 
57.5 MeV and 49.9 MeV, respectively. The fact that one single pion was registered in 
each of the two last (S$_5$) 
detectors shows that the choice of telescope thicknesses was 
reasonable. 
$ $\\

\begin{table}[h]\caption{\label{tab1}\small Telescope geometry (th = 
thickness) together with the proton and pion energy bins (T$_p$, T$_{\pi}$) for the 
telescopes.}\vspace*{2mm}
\begin{tabular}{c|cccc|cccc}
\hline 
 & \multicolumn{4}{|c|}{$ $ 90$^\circ$ telescope} 
& \multicolumn{4}{c}{$ $ 30$^\circ$ telescope}\\ 
\hline  
det. & th & area & T$_p$ & T$_{\pi}$& th & area & T$_p$ & T$_{\pi}$ \\
& (mm) & (mm$^2$) & (MeV) & (MeV) & (mm) & (mm$^2$) & (MeV) & (MeV)\\ 
\hline
S$_1$ & 3 & 60$\cdot$60 & 0 - 17.2 & 0 - 8.7 & 3 & 60$\cdot$60 & 0 - 16.8 & 0 - 
8.4 \\  S$_2$ & 5 & 85$\cdot$85 & 17.2 - 29.1 & 8.7 - 13.8 & 5 & 70$\cdot$70 & 
16.8 - 28.9 & 8.4 - 13.3 \\  S$_3$ & 20 & 90$\cdot$90 & 29.1 - 57.7 & 13.8 - 
26.2 & 10 & 80$\cdot$80 & 28.9 - 45.2 & 13.3 - 20.4 \\  S$_4$ & 30 & 95$\cdot$95 
& 57.7 - 86.4 & 26.2 - 38.8 & 20 & 90$\cdot$90 & 45.2 - 68.3 & 20.4 - 30.2 \\ 
S$_5$ & 50 & 100$\cdot$100 & 86.4 - 122.6 & 38.8 - 57.5 & 50 & 100$\cdot$100 & 
68.3 - 109.2 & 30.2 - 49.9 \\  
\hline
\end{tabular}
\end{table}

$ $\\

\subsection{Electronics and data acquisition}

In order to identify $\pi^+$, the PMT anode 
signal of each detector was split and sent to two independent charge integrating ADCs with 
individually adjustable gates. The "prompt-gated" ADC had a 100 ns long gate which was 
open for integration 5~ns before the analogue PMT signal appeared. The "delay-gated" ADC, 
also had a 100 ns long gate, which was opened when the largest amplitude of the analogue 
pulse was reached. This opening time was individually adjusted from the proton 
pulses, easy to recognize with beam on target.

For each master trigger which required signals in the first two detector elements, 
S1*S2, the hit-pattern was registered and stored in ROOT-tree files together with all 
the ADC and TDC information. A local, VME based data acquisition system was used for 
this storage.

\subsection{Pion identification}

Figure 2 presents typical on-line data. The upper plot shows the prompt 
$\Delta$E - E correlation (in this case, S$_2$ vs S$_3$ in the 
30$^\circ$ telescope) after the ADC and TDC patterns have been used to 
identify the detector element in which the pion stopped. No other constraints 
have been imposed. The existence of three groups of particles - 
protons, pions and e-$\gamma$ background - is obvious. It is important to note 
that the PMT gains were set so that all protons were registered (see 
section 2.6). 

\begin{figure}
\begin{center}
\includegraphics[angle=0, width=0.6\textwidth]{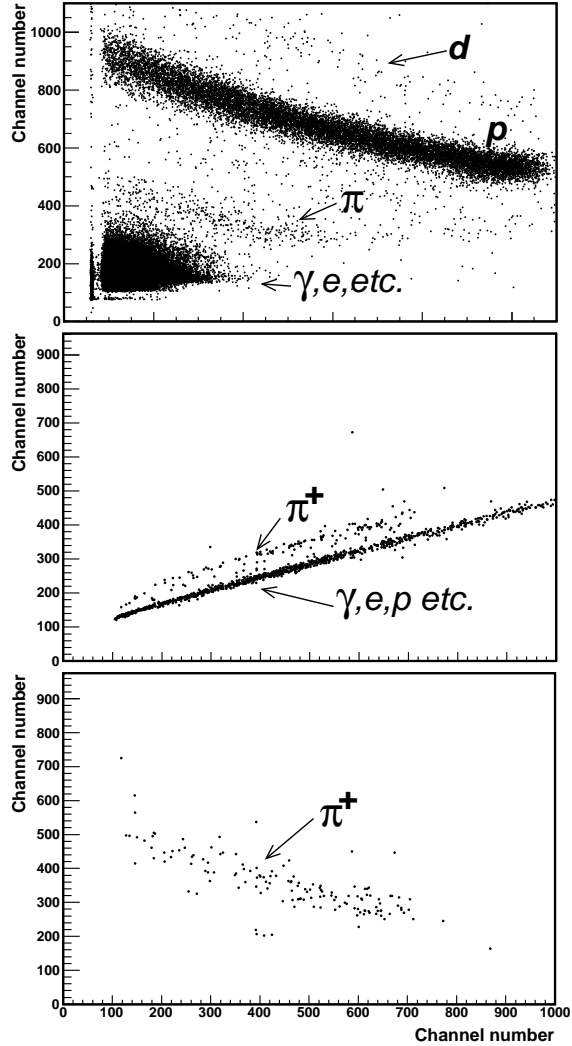}
\caption{(Upper) Prompt $\Delta$E - E plot for 
particles stopping in the S$_3$ detector of the 90$^\circ$ telescope. (Mid) 
Delayed vs. prompt ADC signals for events lying in the pion band in the upper plot. 
(Lower) Prompt $\Delta$E - E plot revisited, now for those events defined as $\pi^+$ 
in the middle plot.} \end{center}
\vspace{0.3in}
\end{figure}

The pion band contains both $\pi^+$ and $\pi^-$. $\pi^+$ were then selected 
in the E$_{prompt}$ vs. E$_{delay}$ plots for the "stop" detector 
(Fig. 2 - mid). The muon from the $\pi^+ \rightarrow \mu^+ \nu$ decay 
(nearly 100$\%$ branching ratio) can be detected with a certain efficiency 
(see section 2.5). The $\pi^-$ is normally absorbed by a carbon atom of the scintillator 
and it thus appears in the E$_{prompt}$ - E$_{delay}$ plot either in the lower band if 
only neutral particles are emitted from the disintegrating carbon nucleus or in a random background 
position if charged particles are emitted. Selecting the upper band in Fig. 2 
(mid) thus provides clean $\pi^+$ identification as evidenced by Fig. 2 (lower), 
which shows only those events originally present in Fig. 2 (upper) which pass the 
$\pi^+$ identification.

Pion-energy intervals were defined from the energy-range intervals that stopped pions 
have in the detectors (Table 2). Polynomial fits, to the energy distributions, 
using the detector half thicknesses were then used to determine the final energies. 
This resulted in systematic uncertainties of $\sim\pm$1 MeV in the pion energies. 
Note that linear scintillator response functions have been used when pion detection 
thresholds are artificially introduced in software.  
 
\subsection{Determination of the $\pi^+$ detection efficiency}

The number of $\pi^+$ obtained in the three-step analysis process described 
above was corrected for the following processes:\\
i)  pion decay in flight.  \\
ii) nuclear reactions in the target and detector material. \\
iii) sliding trajectories (detector geometry). \\
iv) $\mu^+$ detection efficiency.\\

The first three processes make pions undetectable, while correction iv) 
accounts for losses of $\pi^+$ in the E$_{prompt}$ - E$_{delay}$ plot when 
muons leave the stop detector before decaying ($\mu^+ \rightarrow \e^+ \nu \bar{\nu}$). 
In the efficiency calculations, it was assumed that half of the muon energy 
(0.5 $\cdot$ 4.2 MeV) must be deposited in the detector for the pion to appear in 
the proper band in the E$_{prompt}$ - E$_{delay}$ plot. All effects have been calculated 
analytically following the procedure in Ref.~\cite{Ber} and also by 
Monte-Carlo simulations using GEANT4~\cite{GEA}. Table 2 shows the results.  
Note that the analytically determined, total corrections (ranging from 24 to 
44$\%$) are systematically lower than the results from the Monte-Carlo 
simulations (32 - 69$\%$). This difference is interpreted to result from 
approximations in the analytic scattering kinematics calculations. 
The larger correction factors resulting from the simulations were thus used to obtain the 
true number of photopions. The differences in the corrections were the main 
contributions to the systematic uncertainties in the detection efficiency (see section 2.7).

\begin{table}[h]\caption{\label{tab2}\small Total correction factors for 
detector efficiency in the elements S$_2$ to S$_5$ calculated analytically and simulated with 
GEANT4 (see text for details).}\vspace*{2mm}\begin{tabular}{c|cc|cc}    
\hline 
 &\multicolumn{2}{|c|}{$\alpha$ (90$^0$) telescope} & \multicolumn{2}{c}{$\beta$ (30$^0$) telescope} \\
\hline  
Det. & Analytic corr. & M-C corr. & Analytic corr. & M-C corr.\\
\hline
S$_2$ & 1.44 & 1.69 & 1.44 & 1.69\\  
S$_3$ & 1.31 & 1.48 & 1.33 & 1.54\\  
S$_4$ & 1.28 & 1.39 & 1.30 & 1.42\\  
S$_5$ & 1.24 & 1.32 & 1.26 & 1.37\\  
\hline
\end{tabular}
\end{table}

\subsection{Normalization and average beam energy}

Because of the low data acquisition rate, $\ll$ 1$\cdot$10$^3$ events/s,  
no proton rejection was necessary. Consequently, the registered protons could be 
used for absolute normalization. A mixture of inclusive and semi-exclusive ($\gamma$,$p$) 
cross-sections have been utilized for this normalization 
~\cite{Cro,Ter,Mat,Ang,Rui}. 

The fundamental, triple differential cross-section 
that was extracted from this experiment is,

\begin{equation}
\frac{d^3\sigma_{\pi,p}}{dE_{\gamma}d\Omega_{\pi,p} dT_{\pi,p}}(E_{\gamma},\Theta_{\pi,p},T_{\pi,p})=
\frac{1}{C_{\pi,p}}\cdot\frac{1}{\Phi_o(E_{\gamma})}
\cdot\frac{N_{\pi,p}(E_{\gamma},\Theta_{\pi,p},T_{\pi,p})}{\Delta E_{\gamma}
\Delta \Omega\Delta T_{\pi,p}},
\end{equation}

where T$_{\pi,p}$ and $\Theta_{\pi,p}$ are the kinetic energy and emission angle of 
protons and pions respectively. C$_{\pi,p}$ are the normalization constants 
(which depend on target mass number, thickness and density) and 
$\Phi_o$(E$_{\gamma}$) is the bremsstrahlung spectrum of incident photon energies. 
N$_{\pi,p}$ is the efficiency-corrected pion(proton) yield. $\Delta \Omega$ is the solid angle 
subtended by the detector. The most extensive ($\gamma$,$p$) data were found for 
photon collisions on $^{12}$C. Total, inclusive cross-sections could in this case 
be found at energies $\ge$ 200 MeV~\cite{Cro} while total cross-sections at lower energies  
must be estimated from missing energy biased data~\cite{Ter}.  
The normalization constant C$_{p}$ was obtained from the number of 
protons registered in all of the stop detectors, {\em$\sum_{E_{\gamma}=0}^{189}
\sum_{T_p = 17}^{T_{p,max}}N_p(E_{\gamma},\Theta_{p},T_p)$} and the empirical triple 
differential cross-section integrated over the proper intervals in E$_{\gamma}$ 
and T$_p$. In practice, a polynomial fit to the function f(E$_{\gamma}) = 
\int_{0}^{189}\Phi_o(E_{\gamma})\cdot\int_{17}^{T_{p,max}}\frac{d^3\sigma}
{dE_{\gamma}d\Omega_{p} dT_{p}}(E_{\gamma},\Theta_{p},T_{p}) dE_{\gamma}dT_p$ was extracted.

If the pion triple differential cross-section had been known, C$_{\pi} = k\cdot C_{p}$ 
could be determined in a similar manner (and ideally k = 1). Instead, an ansatz was made 
by extrapolating a polynomial fit to the Fissum et al. data ~\cite{Fissum} (photon 
energy region 184 - 213 MeV) to an energy of 160 MeV, below which the pion 
contribution was neglected. C$_{\pi}$ was then 
calculated in the same way as C$_{p}$. Thus, a polynomial function was fitted to g(E$_{\gamma}) = 
\int_{0}^{189}\Phi_o(E_{\gamma})\cdot\int_{17}^{T_{\pi,max}}\frac{d^3\sigma}{dE_{
\gamma}d\Omega_{\pi} dT_{\pi}}(E_{\gamma},\Theta_{\pi},T_{\pi})dE_{\gamma}dT_{\pi}$. 
The number of pions includes here only those with T$_{\pi} >$ 17 MeV, efficiency corrected accordingly (Table 1). 
As mentioned before (section 2.4), linear scintillator response functions were 
assumed when fractions of "stop" detector energy bins were introduced. 
In a typical example of this procedure the determination of absolute differential 
cross-sections for  $\gamma + ^{12}C$ at 90$^\circ$ k was determined to 0.79 with an uncertainty of 
15$\%$ (see section 2.7). 

The same procedure was carried out for the $\gamma$ 
+ $^{27}$Al and $\gamma$ + $^{2}$H reactions. Because of the limited number of 
"normalizing" data, available for $\gamma$ + $^{27}$Al, also $\gamma$ + $^{40}$Ar data 
were used to guide the A$_T$ dependence of C$_{\pi}$.
   
The ansatz for the g(E$_{\gamma}$) function resulted in 
$<E_{\gamma}>$ = 178$\pm$2 MeV for pion producing events. Introducing the 
data from this work at 178 MeV photon energy, resulted in a decrease of 
$< E_{\gamma} >$ to 176$\pm$2 MeV. Further recursive iterations along 
this line changed the value by $<$ 1 MeV. The $<E_{\gamma}>$ was consequently fixed at 
176$\pm$2 MeV. Within the limits of uncertainty, the average photon energy was 
the same for all three $\gamma$ + $^{27}$Al, $^{12}$C,$^{2}$H reactions. 

\subsection{Systematic and statistical uncertainties}

Statistical uncertainties only are presented with the data in Figs. 3 - 7 with two 
exceptions -  the two upper points (triangles) in Fig. 4. These contain the yield 
of pions with energy below the detector cutoff, which has been extrapolated 
both from data and RELDIS calculations. The extrapolated yields 
contribute with 30$\%$ and 50$\%$ of the differential cross-sections at 
30$^\circ$ and 90$^\circ$ lab angle respectively. The uncertainties of these yields, 
which are introduced in the error bars of the two points in Fig. 4, 
are set to 30 $\%$, based on the difference between the empirical and 
RELDIS extrapolations. 

Three major sources of systematic uncertainties in eq. (1) were identified. First, 
the systematic uncertainty in the detector efficiency has been taken as the 
difference in the results obtained from the analytic and Monte-Carlo calculations. 
This gives a pion-energy dependent uncertainty ranging from 7$\%$ at high energy to 
17$\%$ at low energy (Table 2). The second source of uncertainty comes from the 
manner in which the cuts identifying the pion yield was applied. This 
uncertainty was estimated to be $\sim$ 9$\%$. The third systematic uncertainty enters via 
C$_{p}$ through its dependence on the ($\gamma$,$p$) cross-section data taken from 
the literature. By comparing the pion yields from the $\gamma$ + $^{12}$C reaction that are obtained 
with and without the use of the data from Ref. ~\cite{Cro} to estimate C$_{p}$, this uncertainty 
was determined to be 12$\%$. Other sources of uncertainty were negligible. This 
resulted in total systematic uncertainties for the $\gamma$ + $^{12}$C 
reaction that range from 17$\%$ at the highest pion energy to 23$\%$ at the lowest. 
The uncertainties are somewhat larger (up to 28$\%$) for the 
other two reactions. 

\section{The RELDIS model for photonuclear reactions}
\label{RELDIS}

Below pion production threshold, at E$_{\gamma} \sim 140$ MeV, the de Broglie 
wavelength $\lambda$ is comparable to the distance between nucleons in nuclei 
and photon absorption by a quasi-deuteron is the main 
reaction mechanism. At higher energies (as $\lambda$ becomes comparable to the 
nucleon radius), photons interact mainly with single nucleons, thereby exciting 
baryon resonances and producing mesons. The RELDIS model takes into account these 
two competing channels. Here, the input to the code relevant for photon energies 
close to the pion production threshold is described.  
More details can be found in Refs.~\cite{Iljinov,Pshenich}.

\subsection{Calculation of quasi-deuteron absorption}
\label{QD}

The two-nucleon photoabsorption cross-section on a heavy nucleus 
$\sigma^{QD}_{\gamma A}$ is taken from the quasi-deuteron 
model of Levinger~\cite{Levinger}, as modified in Ref.~\cite{Lepretre}, 

\begin{equation}
\sigma^{QD}_{\gamma A} = k Z (1-Z/A) \sigma_d^{exch}.
\label{eq:QD}
\end{equation}

Here $\sigma_d^{exch}$ is the meson-exchange part of the cross-section 
$\sigma_d$ for deuteron photodisintegration, $\gamma d \rightarrow 
np$~\cite{Laget:1978}, $A$ and $Z$ are the mass and charge numbers of the 
target nucleus and $k\approx$ 11 is an empirical constant taken from the 
analysis of Ref.~\cite{Lepretre}.  

The cross-section $\sigma^{QD}_{\gamma A}$ decreases strongly 
with photon energy. Nevertheless, the two-nucleon absorption mechanism competes 
noticeably with the single-nucleon absorption up to $E_\gamma \sim$ 0.5 GeV.

\subsection{Simulation of pion photoproduction on nucleons in a nucleus}
\label{SinglePion}

The main single-nucleon photoabsorption mechanism  is meson production. 
The two-body channel $\gamma N \rightarrow \pi N$ dominates up 
to $E_\gamma \sim$~0.5 GeV. This process was calculated in the 
framework of the phenomenological approach of Ref.~\cite{Walker,Metcalf}.
In this approach, the pion photoproduction amplitude contains  
Breit-Wigner resonant terms, Born terms and a weakly energy-dependent 
``background'' contribution. The latter was used as an adjustable parameter. 
Masses and widths of the resonances were taken from $\pi$N-scattering data, with 
 their amplitudes taken as free parameters. The excitation of six different 
baryon resonances were considered with the $\Delta$(1232), $N^\star$(1520) and 
$N^\star$(1680) resonances the most important of them.

Tables of total reaction cross-section data together with photopion angular 
distributions calculated according to
Refs.~\cite{Walker,Metcalf} were taken from Ref.~\cite{Corvisiero}. 

Both the total and the differential cross-sections of the $\gamma p \rightarrow 
\pi^+ n$, $\gamma n \rightarrow \pi^- p$ and $\gamma p \rightarrow \pi^0 p$ 
processes are well described by the $\gamma N$ Monte Carlo event 
generator~\cite{Iljinov} used by the RELDIS model. Presently, the 
$\gamma N$ event generator used in the RELDIS model extends the event generator 
by Corvisiero et al.~\cite{Corvisiero} to higher photon energies (up to 10 GeV) and 
multiple pion production, as demonstrated in Refs.~\cite{Iljinov,Pshenich}.

\subsection{Secondary interactions of photohadrons}

Hadrons produced in a primary $\gamma N$ or $\gamma d \rightarrow np$  
interaction initiate a cascade of successive hadron-nucleon
collisions inside the target nucleus during the cascade
stage of the photonuclear reaction. 
Calculations are based on a Monte Carlo technique used 
to solve the equation that describes hadron transport 
in the nuclear medium. The target nucleus is considered to be a mixture
of degenerate Fermi gases of neutrons and protons in a spherical
potential well with a diffuse boundary. 
By using the effective real potentials for nucleons and pions, the influence of
the nuclear medium on cascade particles is taken into
account. It should be stressed that this potential is taken to be constant 
although it should depend on the nuclear density as well as on the pion kinetic 
energy. However, it is beyond the scope of the present experiment to 
provide the very precise data necessary to tune these dependences. 

The momentum distributions of nucleons in the nuclei are
calculated in the local-density approximation of the Fermi-gas
model. The distribution of nuclear density is approximated by a set
of step functions for the nuclear radius. The Coulomb potentials for charged 
cascade particles are calculated for each density zone.
 
The cross-sections for pion interactions such as 
$\pi N\rightarrow \pi N$,  $\pi (N N)\rightarrow N N$,  $\pi N\rightarrow \pi\pi 
N$ etc, as well as nucleon-induced processes, $NN\rightarrow NN$, $NN\rightarrow 
\pi NN$, ,$\ldots$ etc in the nuclear medium are assumed to be the same as in 
vacuum, except that the Pauli principle prohibits the transition of the
cascade nucleons into states already occupied.

\section{Experimental data and comparison to RELDIS calculations}

Data on $\pi^+$ cross-sections were measured in the T$_{\pi}$ interval(s) 8.7(8.4) - 57.5(49.9) MeV. 
No pions with higher energies are expected (section 2.2). For simplicity, these pions 
are subsequently denoted as T$_{\pi} > $ 9 MeV pions. 
Due to limited statistics, data points are presented one per "stop" detector  
(Table 1), all at beam energy E$_{\gamma}$ = 176 ($\pm$ 2) MeV (recall section 2.6). 

In the near-threshold RELDIS simulations used for comparison to 
the data in Figs. 3 - 7, the only process for $\pi^+$ production is 
$\gamma p \rightarrow \pi^+ n$. Photopions can then face 
elastic scattering from nucleons in the target nucleus, charge-exchange reactions 
(such as $\pi^+ n \rightarrow \pi^0 p$) or absorption on two nucleons. The key parameter in these 
simulations is the real part of the pion potential $V_{\pi}$ which is assumed to 
be attractive, $V_{\pi}<$ 0. This potential is taken as an empirical parameter, 
independent of the pion energy. Its value has been derived from pion production 
data reasonably close to threshold. The imaginary part of the pion potential is simulated 
by taking into account the  pion absorption reaction $\pi (N N)\rightarrow N N$.

Because of the limited statistics and large systematic uncertainties, it is 
difficult to compare the double differential cross-section data to the 
predictions from RELDIS, which themselves are very sensitive to 
small changes in the input parameters. Therefore, total (energy and angle integrated) 
cross-section data are presented and discussed first. Least squares fits to Boltzmann functions (energy) and polynomial 
functions (angular) have been introduced to carry out these integrations.  

\subsection{Total yield of $\pi^+$}

Data from Ref.~\cite{Fissum} demonstrate a falloff of 
$\sigma_{\pi^+}$ with decreasing E$_{\gamma}$ towards the effective reaction threshold. 
The measured total cross-section data  for pions with T$_{\pi} >$ 17 MeV 
(the cutoff in Ref.~\cite{Fissum}) from this experiment 
and from selected parts of Ref. ~\cite{Fissum} are shown in Table 3. The average systematic 
uncerainty in the $^{1}$H and $^{12}$C data from Ref. ~\cite{Fissum} was $\sim$ 17$\%$, i.e. somewhat less 
than in the present experiment (see Table 3).

\begin{table}[h]\caption{\label{tab3}\small Total cross-section data for $\pi^+$ photoproduction 
with T$_{\pi} >$ 17 MeV from this work and from Ref. ~\cite{Fissum}. Statistical uncertainties 
are shown.}\vspace*{2mm}\begin{tabular}{cccccc}    
\hline 
 & & & $\sigma (\mu b)$ & & \\
 \hline
 E$_{\gamma}$ (MeV)& $^{1}$H & $^{2}$H & $^{12}$C & $^{27}$Al & $^{40}$Ca $^a$\\   
\hline
176 (this work) & - & 0.8$\pm$0.2 $^b$ & 7.2$\pm$1.2 $^c$ & 43$\pm$7 $^d$ & - \\
184 ~\cite{Fissum} & - & - & 31.1$\pm$2.0 & - & 131.8$\pm$7.8\\  
194 ~\cite{Fissum} & 82.0$\pm$3.7 & - & 64.4$\pm$2.8 & - &177.8$\pm$11.7\\  
204 ~\cite{Fissum} & 92.3$\pm$4.2 & - & 107.8$\pm$3.8 & - & 354.3$\pm$13.0\\  
213 ~\cite{Fissum}  & 103.7$\pm$4.7 & - & 160.7$\pm$4.5 & - & 498.1$\pm$15.0\\  
\hline \end{tabular} \\
$^a$ Systematic error associated with the background 8.8 $\mu$b ~\cite{Fissum} \\
$^b$ Systematic error 0.2 $\mu$b, $^c$ Systematic error 1.4 $\mu$b, $^d$ Systematic error 12 $\mu$b  \\
\end{table}

The process that RELDIS uses to produce pions should result in $\sigma_{\pi^+}$ scaling 
with Z$_{target}$. Thus, $\sigma_{\pi^+}$ / Z$_{target}$ vs. 
E$_{\gamma}$ is plotted in Fig 3. The Fissum et al. data show that heavy nuclei are 
less efficient in photoproducing pions. This may be due to the enhanced reabsorption 
of $\pi^+$. The strongest 
deviation from Z$_{target}$ scaling appears actually in the $\gamma$ + $^1$H cross-section, 
which is 5 - 10 times more efficient in producing $\pi^+$ than the $\gamma$ + 
nucleus reactions~\cite{Fissum}.

\begin{figure}
\begin{center}
\includegraphics[angle=0, width=1.0\textwidth]{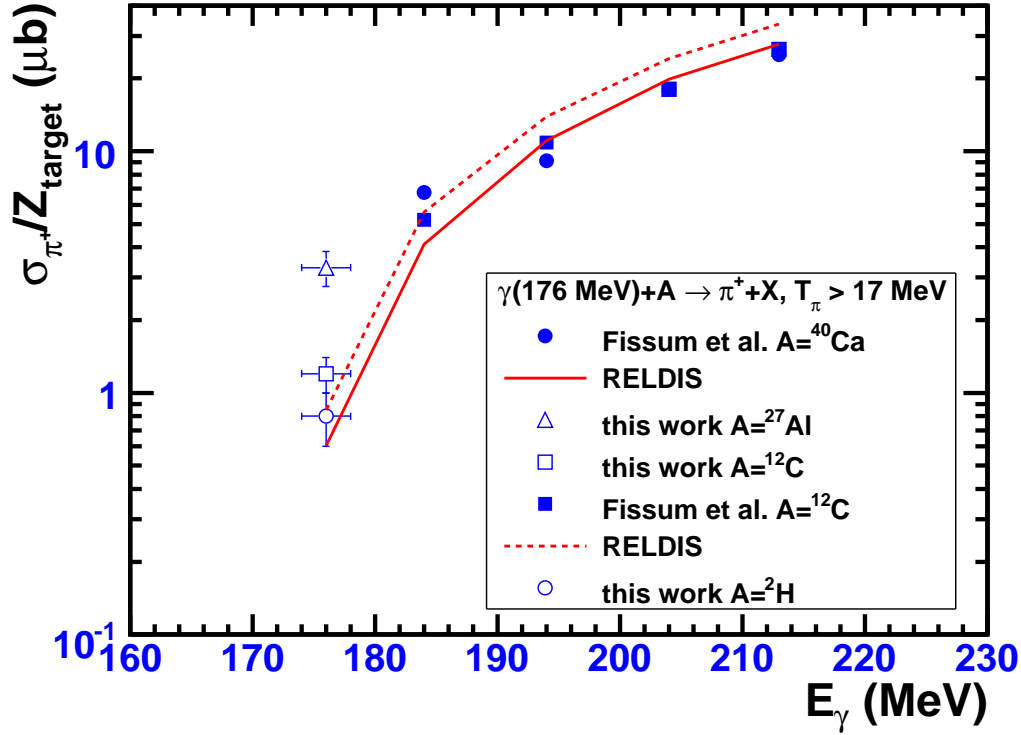}
\caption{Total yield of $\pi^+$ with T$_{\pi} >$ 
17 MeV (the pion-detection threshold from Ref. ~\cite{Fissum}) as a function of 
E$_{\gamma}$. Points represent data (this work and 
Ref.~\cite{Fissum}) according to the legend and curves represent RELDIS 
calculations with a real pion potential of -20 MeV for the $\gamma$ + $^{12}$C 
(solid) and $\gamma$ + $^{40}$Ca (dashed) reactions. Systematic uncertainties 
of the measurements are given in Table 3. See text for details.}
\end{center}
\vspace{0.3in}
\end{figure}

The new data shown in Fig. 3 exhibit an enhanced falloff of pion production in 
the $\gamma$ + $^{12}$C reaction, consistent with the predictions of the 
RELDIS model. There is of course a shift from the ($\gamma,N$) channel to the 
quasi-deuteron absorption channel in this energy region but on the other hand, 
the $\gamma$ + $^{27}$Al data do not show this enhanced falloff. This is 
in contradiction to the RELDIS results and the deviation is well outside 
statistical and systematic uncertainties. RELDIS results also exhibit a steeper falloff 
for heavier nuclei (note that Fig. 3 shows $\gamma$ + $^{40}$Ca calculations). The enhanced 
production in heavier nuclei exhibited by the data taken in this experiment, may be 
due to an underestimation of the tail of the internal momentum distribution of 
the nucleons or a signal that coherent processes should be included in the model. 

Another remark to Fig. 3 is that the $\pi^+$ cross-section in the 
$\gamma$ + $^{2}$H reaction does not follow the elementary $\gamma$ + $p$ trend, but rather 
the trend established by the $\gamma$ + nucleus reactions. It thus appears as if the 
presence of one single pn pair leads to the normal competition between the 
single nucleon and quasi-deuteron processes. 
\begin{figure}
\begin{center}
\includegraphics[angle=0, width=1.0\textwidth]{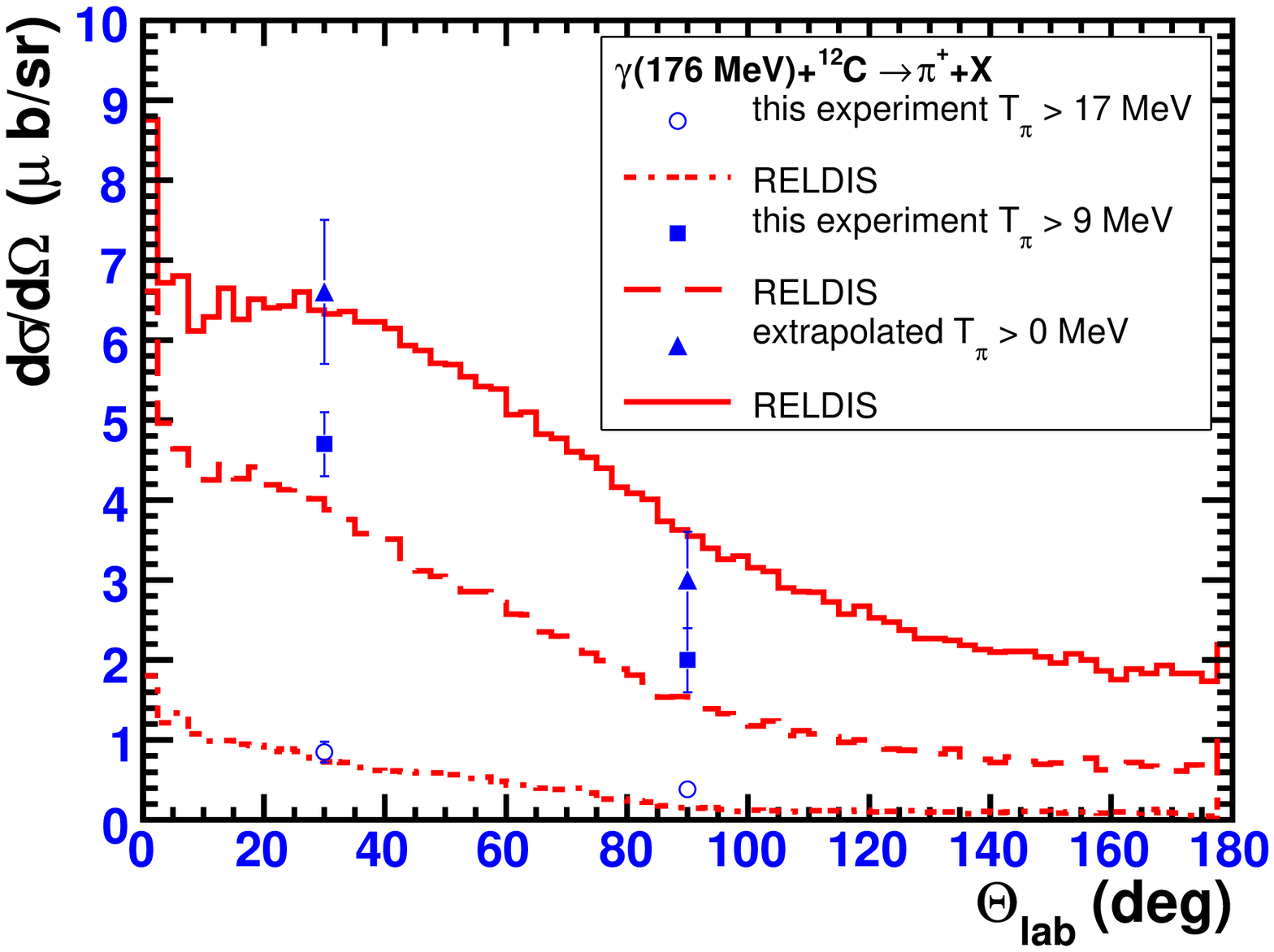}

\caption{Angular distributions of $\pi^+$ with 
T$_{\pi} >$ 17 MeV (open points), $>$ 9 MeV (squares) and $>$ 0 MeV (triangles),
emitted in $\gamma$ + $^{12}$C reactions at 176 MeV. Systematic uncertainties of 
the measurements are between 17 and 23$\%$. Curves represent RELDIS 
calculations with a real pion potential of -20 MeV.} 

\end{center}
\vspace{0.3in}
\end{figure}

It should finally be stressed that in this near-threshold 
energy domain, the available phase-space for pion production becomes increasingly limited and a 
high detection threshold (e.g. 17 MeV at 90$^\circ$) makes this restriction even more important.

\subsection{Angular distributions of produced $\pi^+$}

In Fig. 3, the falloff of $\pi^+$ photoproduction at 
energies $<$ 180 MeV was noted. Fig. 4 shows that the differential cross-section, d$\sigma$/d$\Omega$, 
decreases by a factor 4 - 7 if the detector limit is raised from 9 MeV (this experiment) 
to 17 MeV (detection threshold in Ref. ~\cite{Fissum}). The extrapolated increase in pion 
yield if the detection threshold could be lowered to 0 MeV, is a factor of 1.5 - 2. The trend exhibited by the data 
is reasonably well reproduced by RELDIS, with V$_\pi$ = -20 MeV. Systematic uncertainties 
associated with the data points are are between 17$\%$ and 23$\%$ and the 
angular acceptance is $\pm$ 5.4$^\circ$ and $\pm$ 7.3$^\circ$ for the forward and 90$^\circ$ telescopes 
respectively.

\begin{figure}
\begin{center}
\includegraphics[angle=0, width=1.0\textwidth]{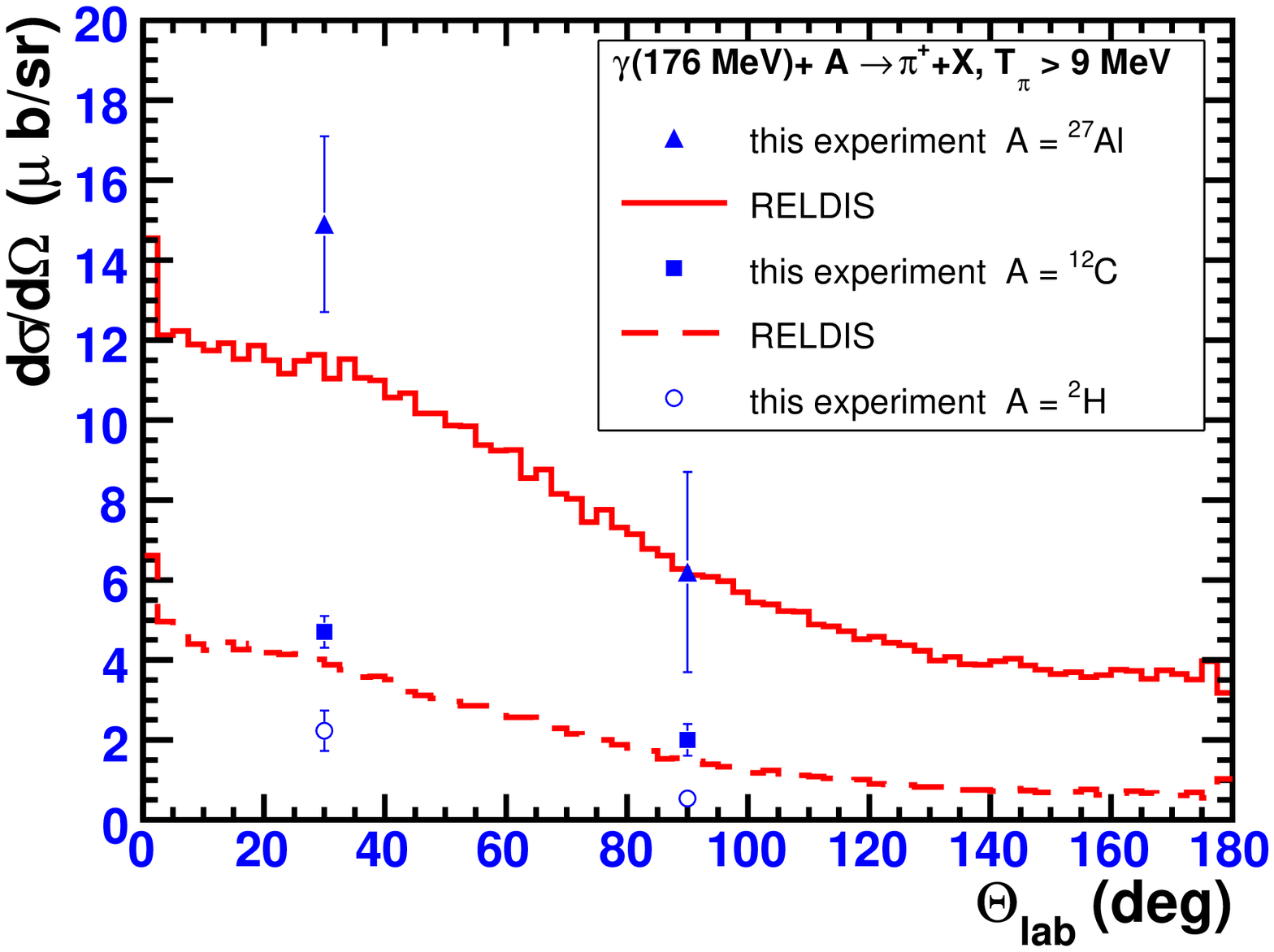}

\caption{Angular distributions of $\pi^+$ with 
T$_{\pi} >$ 9 MeV in  $\gamma$ + A (see legend) reactions at 176 MeV. The systematic uncertainties are 
between 17$\%$ and 28$\%$. The angular acceptance is $\pm$ 5.4$^\circ$ and $\pm$ 7.3$^\circ$ 
for the 30$^\circ$ and 90$^\circ$ respectively. Curves 
represent RELDIS calculations for A = $^{12}$C (dashed) and A = $^{27}$Al 
(solid) with a real pion potential of -20 MeV. } 
\end{center}
\vspace{0.3in}
\end{figure}

\begin{figure}
\begin{center}
\includegraphics[angle=0, width=1.0\textwidth]{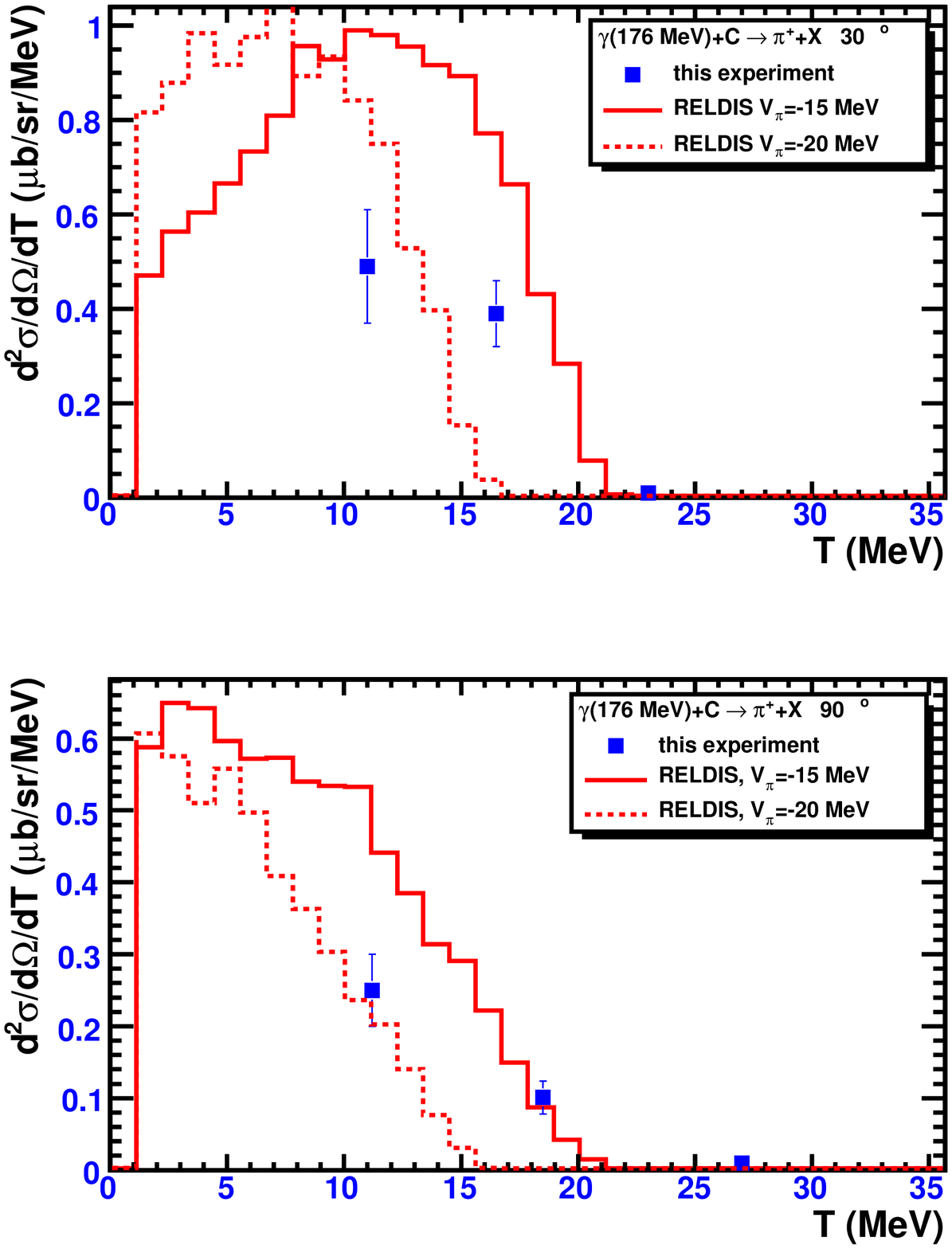}

\caption{Double-differential cross-section data for 
$\pi^+$ photoproduction with T$_{\pi} >$ 9 MeV in  $\gamma$ + $^{12}$C reactions at 176 MeV. 
The detector angle is 30$^\circ$ (upper panel) and 90$^\circ$ (lower panel).  
Note the experimental points at T$_{\pi} >$ 20 MeV. Systematic uncertainties are between 20$\%$ and 25$\%$
and the uncertainty of the energy positions typically 1 MeV. Curves represent RELDIS 
calculations with a real pion potential of -20 MeV (dashed) or -15 MeV (solid).}   \end{center}
\vspace{0.3in}
\end{figure}

This is also essentially true for the $\gamma$ + $^{27}$Al reaction, as shown 
in Fig. 5. The excess pion yield was here essentially found in the forward hemisphere. 
This fact goes hand-in-hand 
with the assumption of an extended internal momentum distribution since the 
coupling of the relative momentum vector to the internal momentum vector 
produces a forward-focusing effect. However, there are other tentative 
explanations such as too strong reabsorption in the model that are plausible. 
Finally, we note again that the $\gamma$ + $^{2}$H   
reaction behaves more like the reaction on heavier nuclei than the elementary process. 

\begin{figure}
\begin{center}
\includegraphics[angle=0, width=1.0\textwidth]{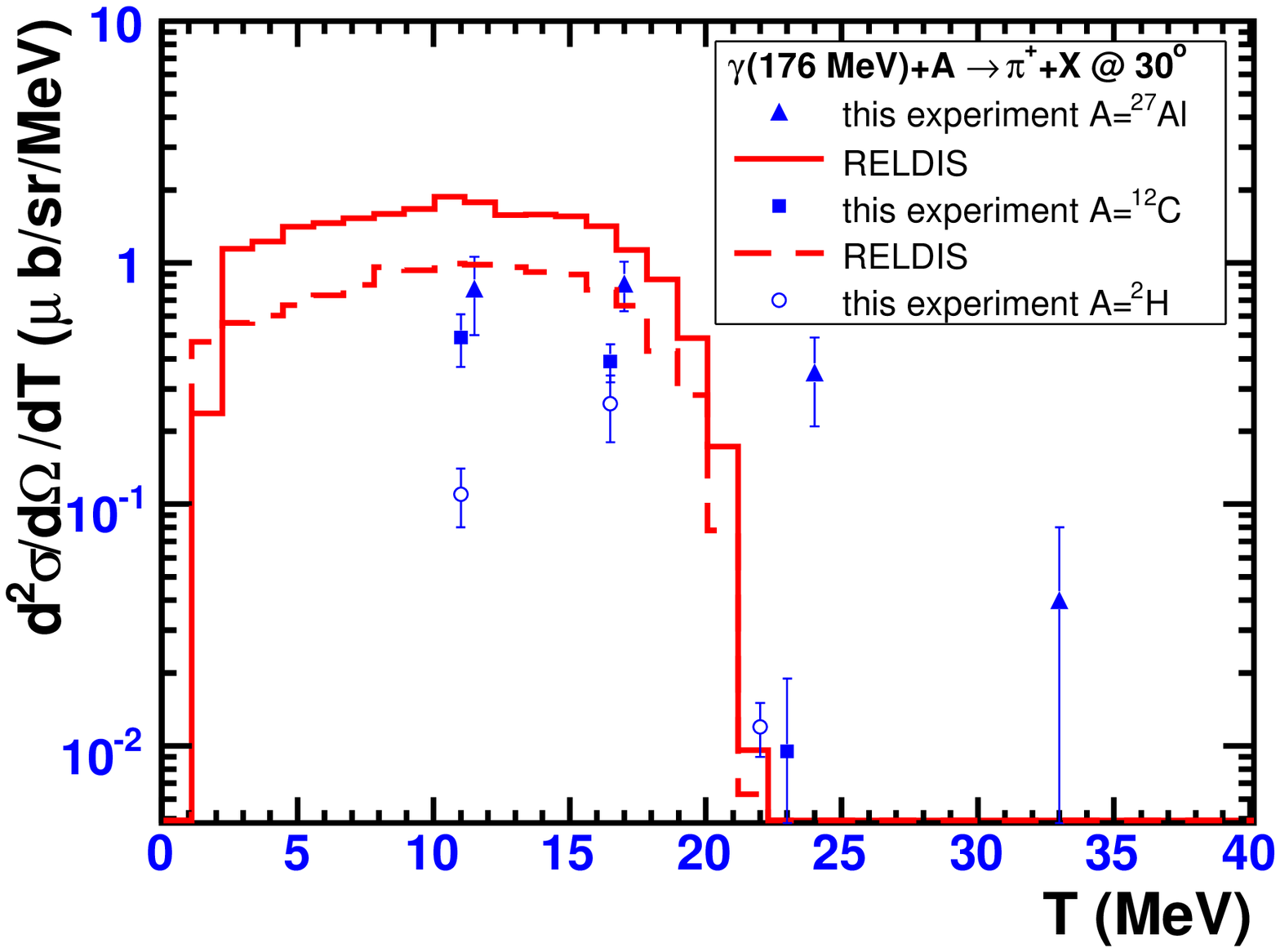}

\caption{Double-differential cross-section data for $\pi^+$ photoproduction 
detected at 30$^\circ$, with T$_{\pi} >$ 9 MeV in  $\gamma$ + $^{2}$H,$^{12}$C 
and $^{27}$Al (see legend) reactions at 176 MeV. Systematic uncertainties 
are between 22$\%$ and 28$\%$ and the uncertainy in energy position typically 1 MeV. 
Curves represent RELDIS calculations for Al and C reactions with a real pion potential of -20 MeV.}    
\ 
\end{center}
\vspace{0.3in}
\end{figure}

\subsection{Double-differential cross-section}

The conclusion from the d$\sigma$/d$\Omega$ data of $\pi^+$ in $\gamma$ + 
$^{12}$C reactions (Fig. 4) was a slight underprediction by RELDIS with V$_{\pi}$ 
= -20 MeV which, however, rather tends to become an overprediction when 
extrapolating the data to T$_{\pi}$ = 0 MeV. The behaviour of the double differential 
cross-section as a function of pion kinetic 
energy (Fig. 6) at 30$^\circ$ and 90$^\circ$ makes the overestimation 
even more plausible. The obvious overprediction for low T$_{\pi}$ is to some 
extent compensated by more extended high energy tails in the data. 

The same tendencies are more pronounced for the $\gamma$ + $^{27}$Al data (Fig. 
7). In this case, RELDIS calculations exhibit no high-energy tail and 
the expected maximum in a Coulomb shifted spectrum of $\pi^+$ from the 
one-nucleon scattering process seems to fall at lower energies in the 
calculations. All calculated curves in Fig. 6 and 7 have a steep rise in the 
second bin, corresponding to an effective Coulomb barrier of $\sim$1 MeV. This 
appears to be too low, at least for the $^{27}$Al case, where a more realistic 
Coulomb shift of 3 - 4 MeV would improve the comparison. The choice of the optimal 
pion potential may suffer from this mismatch. More data, especially on reactions with 
heavy targets, are however needed to set this question. At present it appears 
reasonable to believe that the potential falls in the interval -20 MeV to -15 MeV for light 
nuclei, with the lower value more representing $^{27}$Al and the higher (-15 
MeV) representing $^{12}$C. 

The form of the 30$^\circ$ energy spectrum of $\pi^+$ 
from $\gamma$ + $^{2}$H reactions (Fig. 7) is somewhat different in that the low 
energy point ($\sim$11 MeV) has a low d$^2\sigma$/d$\Omega$dT value. Even if 
this fact is significant, it must be stressed that the H$_2$ data are derived as 
a difference between the CH$_2$ target and C target data, which makes the systematic 
uncertainties larger, 26 - 28$\%$, than for the other targets. 

\section{Conclusions}

Near-threshold ($\gamma$,$\pi$) reactions in light nuclei may be described by an intranuclear 
cascade model like RELDIS. This means that the basic 
processes in RELDIS at these energies namely quasi-deuteron absorption and photon absorption on a single 
bound nucleon, should be dominant. Chosing depth of the real pion potential of -20 to -15 MeV 
results in RELDIS predictions which match the data. Our approach to pion-nucleus reactions is similar to the one adopted in 
the Liege intranuclear cascade model~\cite{Aoust}, which was recently 
reconsidered in theoretical investigations of the average potential energy felt 
by a pion inside the nucleus. Our finding of $V_{\pi}$ = -20 MeV agrees well with 
the values found in the literature \cite{Aoust}, with a surface component of 
$V_{\pi}$ between -5 and -25 MeV. Since low-energy pions, produced in the 
central part of a nucleus, are more easily absorbed than those produced close to 
the nuclear surface, it is relevant to use this surface component of 
$V_{\pi}$ for the comparison. 

The shapes of the experimental $\pi^+$ angular and (in particular) energy distributions differ 
from RELDIS predictions. Furthermore, the total $\pi^+$ yield in heavier nuclei 
is underestimated by the code by a large factor. Model particulars, such as the effective Coulomb 
potential, the exact form of the degenerate nucleon momentum distribution and the 
energy and density dependence of the pion potential could 
certainly be tuned to improve the comparison. The discrepancies also 
hint that multinucleon channels can play a role not only in 
pion absorption but also in pion production processes   

$\pi^+$ cross-sections per proton are much higher in the elementary $\gamma$ + $p$ 
reaction than in $\gamma$ + nucleus reactions close to threshold. The fact that 
the $\gamma$ + $^{2}$H data follow the $\gamma$ + 
nucleus trend indicates that the balance between quasi-deuteron 
absorption and single-nucleon absorption is achieved already for deuteron reactions. In 
fact, all data from this experiment indicate that the effects of the nuclear 
environment are small for nuclei with A $<$ 30. 

We look forward to studying the details of the pion production mechanism in 
forthcoming tagged ($\gamma$,$\pi$) experiments at MAX-lab at energies up to 
$\sim$220 MeV in the very near future.

\section{Acknowledgements}

The authors acknowledge the outstanding support of the MAX-lab staff 
during this challenging startup period at the upgraded Tagged Photon Facility. 
Funding of this work from the Swedish Research Council, the Knut and Alice Wallenberg Foundation 
and the U.S. Department of Energy is acknowledged.


\end{document}